\documentclass{PoS}

\usepackage[authoryear,square]{natbib}
\bibpunct{(}{)}{;}{a}{}{,}
\usepackage{psfig}
\usepackage{textgreek}

\newenvironment{packed_item}{
\begin{itemize}
  \setlength{\itemsep}{1pt}
  \setlength{\parskip}{0pt}
  \setlength{\parsep}{0pt}
}{\end{itemize}}

\def\HI{H\,{\sc i}}

\def\farcs{\hbox{$.\!\!^{\prime\prime}$}}
\def\la{\ifmmode\stackrel{<}{_{\sim}}\else$\stackrel{<}{_{\sim}}$\fi}
\def\ga{\ifmmode\stackrel{>}{_{\sim}}\else$\stackrel{>}{_{\sim}}$\fi}

\title{Broadband Polarimetry with  the Square Kilometre Array:
A Unique Astrophysical Probe}

\ShortTitle{Broadband Polarimetry with the SKA}

\author{
\speaker{B. M. Gaensler}\thanks{On behalf of the SKA Cosmic Magnetism
Working Group}~\thanks{Present address: Dunlap Institute for Astronomy and
Astrophysics, The University of Toronto}~$^{1,2}$, Iv\'an Agudo$^3$, Takuya Akahori$^{1,4}$, Julie
Banfield$^{5,6}$, Rainer Beck$^7$, Ettore Carretti$^5$, Jamie
Farnes$^{1,2}$, Marijke Haverkorn$^8$, George Heald$^9$, David Jones$^8$,
Thomas Landecker$^{10}$, Sui Ann Mao$^7$, Ray Norris$^{5,2}$, Shane
O'Sullivan$^{1,2,11}$, ~~~~~~Lawrence Rudnick$^{12}$, Dominic
Schnitzeler$^7$, Nicholas Seymour$^{5,13}$, Xiaohui Sun$^1$
\\
$^1$The University of Sydney; 
$^2$ARC Centre of Excellence for All-sky Astrophysics (CAASTRO);
$^3$JIVE; 
$^4$Kagoshima University; $^5$CSIRO; 
$^6$ANU; $^7$MPIfR; $^8$Radboud Universe Nijmegen; 
$^9$ASTRON; $^{10}$DRAO; $^{11}$UNAM;
$^{12}$University of Minnesota; $^{13}$ICRAR -- Curtin University
\\
E-mail: \email{bryan.gaensler@sydney.edu.au}}

\abstract{ 
Faraday rotation of polarised background sources 
is a unique probe of astrophysical magnetic fields in a
diverse range of foreground objects. However, to understand the properties
of the polarised sources themselves and of depolarising phenomena along the
line of sight, we need to complement Faraday rotation data with polarisation
observations over very broad bandwidths.  Just as it is impossible to
properly image a complex source with limited $u$-$v$ coverage, we can only
meaningfully understand the magneto-ionic properties of polarised sources if
we have excellent coverage in $\lambda^2$-space.  We here propose a set of
broadband polarisation surveys with both SKA1 and SKA2, which will
provide a singular set of scientific insights on the ways in which galaxies
and their environments have evolved over cosmic time.
}

\FullConference{
Advancing Astrophysics with the Square Kilometre Array\\
June 8-13, 2014\\
Giardini Naxos, Italy}

\begin{document}

\section{Introduction}

Faraday rotation is a superb probe of astrophysical magnetic fields
at all redshifts. This point was made
strongly in the 2004 Square Kilometre Array (SKA) Science Case, in which \cite{bg04} proposed a
1.4-GHz polarisation survey with the SKA covering 10\,000~deg$^2$, resulting
in a closely spaced ``Faraday rotation measure grid'' of active galactic
nuclei (AGN).  As discussed by \cite{gbf04}, this rotation measure (RM) grid
could then be used to study the detailed magneto-ionic properties of the
Milky Way, nearby galaxies, distant galaxies and galaxy clusters, with the
ultimate goal of
distinguishing between different origins for magnetism throughout
the Universe. ``Cosmic Magnetism'' was correspondingly designated one of the
five Key Science Projects for the SKA \citep{gae04c,cr04}.

A decade later, the relevance and importance of the RM grid have only
increased \citep[see][]{tsg+07,tss09,sab+08,kab+09,gts+10,hrg12,ojr+12,ojg+14,hngm14,ro14,skgt14,agr14}.
Specifically, the RM grid remains a powerful tool for probing foreground
magnetic fields, i.e., cases where the polarised emitting region and Faraday
rotating medium are distinct and well separated, and where the combined
statistical
properties of a large number of sightlines can be used to extract the
global magneto-ionic properties of intervening material \citep{joh14}.

\subsection{Beyond The Rotation Measure Grid}
\label{sec_intro_beyond}

For cases in which the
polarised emission and Faraday rotation occur in the same source, or in
which we are studying the polarisation properties of a single sightline
rather than the large ensemble probed by the RM grid, our understanding of polarisation and Faraday
rotation has substantially deepened and matured over the last ten years.
For example, \cite{bg04} assumed that around
50\% of polarised AGN would suffer from internal
depolarisation, which would manifest itself as a non-linear
dependence of polarisation position angle, $\theta$ on $\lambda^2$ (where
$\lambda$ is the observing wavelength).
However, we now realise that
even quite simple scenarios (e.g., a source consisting of two spatially
unresolved polarised components, each with different RMs) can produce an
apparent linear relationship between $\theta$ and $\lambda^2$, but
corresponding to a spurious value of RM; only when one
considers the fractional polarisation, $\Pi
\equiv \sqrt{(Q^2+U^2)}/I$, as a function of $\lambda^2$ does the observer
realise that something is amiss \citep{frb11}. In such cases, and also in sources for which there is no
linear dependence of $\theta$ on $\lambda^2$, we also now better
appreciate the various ways in which $q \equiv Q/I$, $u \equiv U/I$ and $\Pi$,
all as a function of $\lambda^2$, can
provide detailed information on magnetic fields, ionised gas and turbulence
\citep{bje11,frb11,bfss12,bml12,fgc14,bs14,hf14}. 
Furthermore, it is now
clear that over a fractional bandwidth of 25\% as assumed by
\cite{bg04}, there are considerable degeneracies as to the nature of the
observed Faraday rotation \citep{lgb+11,obr+12}. Finally, while \cite{bg04}
proposed that fitting $\theta$ vs $\lambda^2$ or applying RM synthesis
\cite{bd05} should be sufficient to extract Faraday rotation in most cases,
we now realise that a measurement of RM and its error for a polarised source
is an extremely complex problem that as yet has no optimal solution
\citep[e.g.,][]{frb11,hgnm12,mefj12,ast12,ita+14,kai+14,sra+14}.

\begin{figure}[htb]
\centerline{\psfig{file=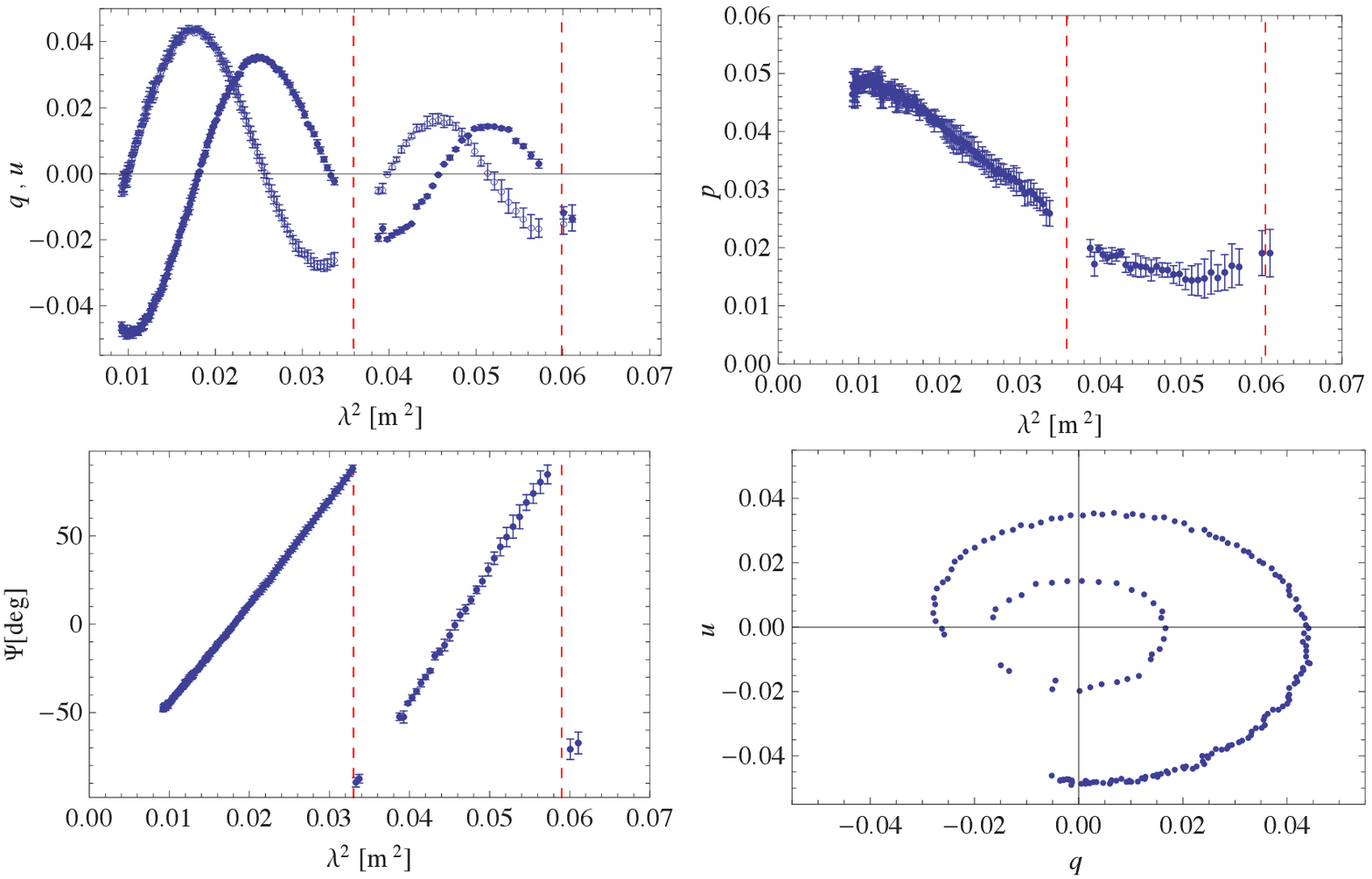,width=0.5\textwidth,clip=}
\psfig{file=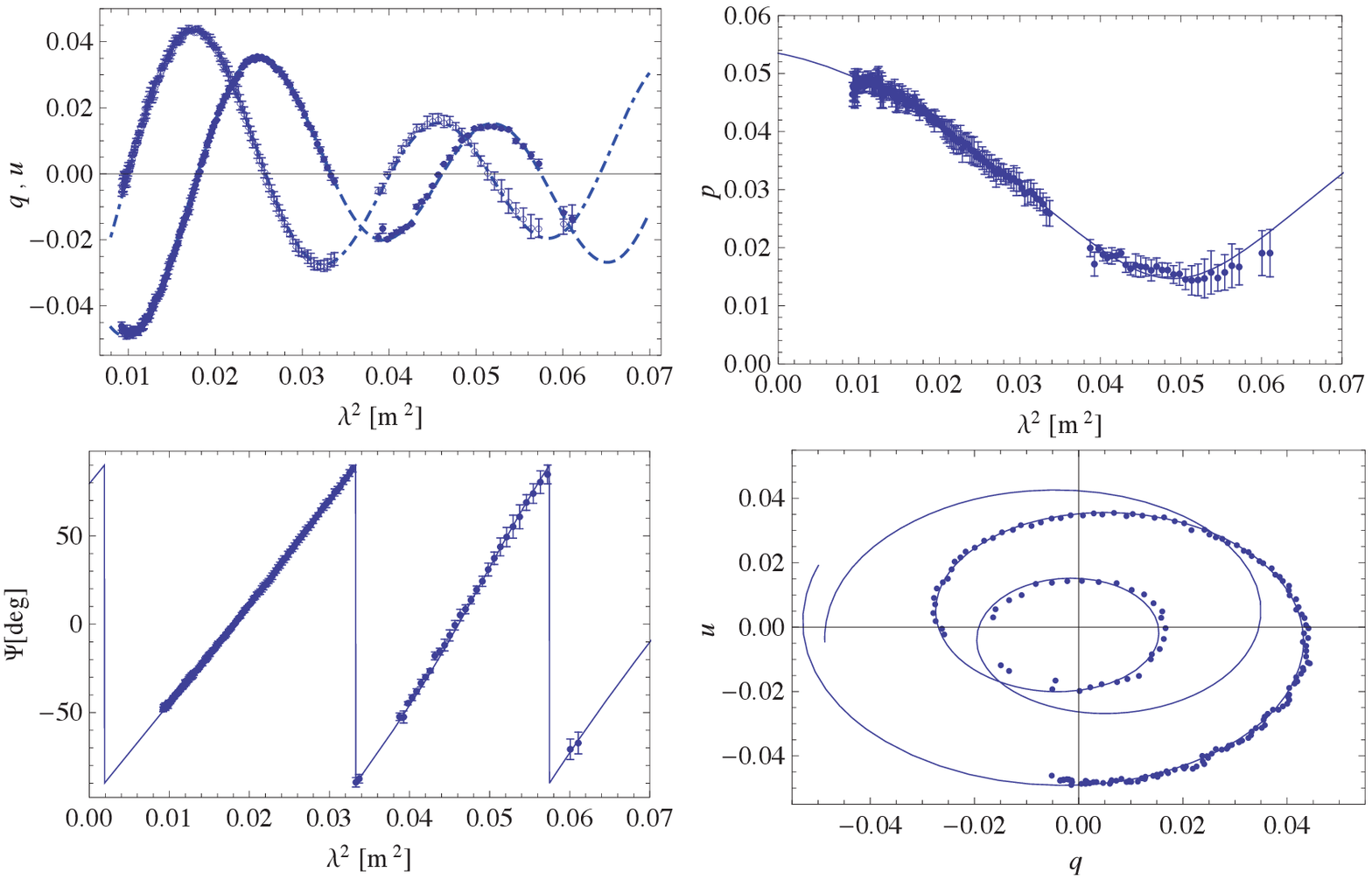,width=0.5\textwidth,clip=}}
\caption{Polarisation data for PKS~B1610--771 over the frequency range 1.1
to 3.1~GHz \citep[adapted from Figure~12 of][]{obr+12}. Top panel: $q$ (open circles)
and $u$ (filled circles) vs $\lambda^2$. On the left-hand side, the vertical
red lines show the range in $\lambda^2$ corresponding
to an observation of bandwidth 350~MHz centred on 1.4~GHz; on the right-hand
side, the data are fit 
over the full bandwidth to a model containing two RM components.
Bottom row: as for the top row, but showing  polarisation angle vs
$\lambda^2$.} 
\label{fig_obr12}
\end{figure}

Some of these issues are exemplified in Figure~\ref{fig_obr12}, which shows
an observation by \cite{obr+12} of the bright polarised quasar
PKS~B1610--771 over the frequency range 1.1 to 3.1~GHz. Three features of
note are apparent from these data:
\begin{enumerate}
\item The plot of polarisation angle vs $\lambda^2$ deviates from a purely
linear dependence;
\item The plots of $q$ and $u$ vs $\lambda^2$ do not show sinusoids of
constant amplitude, and $\Pi$ is not constant as a function of $\lambda^2$;
\item If observed over a relatively narrow bandwidth, none of these
behaviours would be apparent.
\end{enumerate}
For example, \cite{bg04} consider a representative SKA survey with a
bandwidth of 350~MHz centred on 1.4~GHz, corresponding to the
vertical red lines
in the left column of Figure~\ref{fig_obr12}. With such observations, a user would infer
a single foreground RM value of +135~rad~m$^{-2}$, would not realise that
this value was likely in error, 
and would have no capacity to identify or
understand this source's potentially interesting properties.
In contrast, consider the right column of Figure~\ref{fig_obr12}, which
shows the same data but now fit with a simple model consisting
of two spatially resolved components with different RMs.  This
shows an excellent match to the data, and allowed \cite{obr+12} to infer
that this source most likely contains two polarised knots with
RMs\footnote{Note that RM~$= +135$~rad~m$^{-2}$ (as
would be inferred from observations over a narrow bandwidth) is not even
in the range bracketed by these two values.} of +107~rad~m$^{-2}$ and +79~rad~m$^{-2}$. 

\subsection{Broadband Polarimetry with the SKA}

Beyond the simple example described in \S\ref{sec_intro_beyond}, there are a
whole range of complex behaviours that can only be identified and
distinguished if radio polarisation data are recorded over
sufficient bandwidth and at high angular resolution and sensitivity.
Just as it is impossible to properly image a
complex source with limited $u$-$v$ coverage, we can only meaningfully
understand the magneto-ionic properties of the polarised sources themselves
if we have excellent coverage in $\lambda^2$-space.

In this Chapter, we highlight the unique new physical insights
provided by SKA polarisation surveys over a wide contiguous
frequency range. It is important to emphasise that these broadband
experiments complement the RM grid, rather than supersede it. The purpose of
the RM grid is to study magnetic fields in extended foreground sources such
as the Milky Way and the 
intergalactic medium (IGM).\footnote{Sources such as that shown in
Figure~\ref{fig_obr12} do not invalidate the RM-grid approach, because
only $\la25\%$ of polarised sources are
expected to show behaviour that deviates from the idealised ``Faraday thin''
case \citep{lgb+11}. In addition, the wideband data described here will
allow us to explicitly exclude such sources from the RM grid.}
In contrast, broadband data either provide detailed information on {\em
individual}\ sightlines, or can better probe the {\em intrinsic}\
properties of the polarised emitters and their 
environments. The RM grid primarily probes Faraday rotation,
while the broadband studies proposed here primarily probe depolarisation.
More fundamentally, broadband polarimetry transcends
the study solely of magnetic fields, and instead becomes a highly sensitive
tool for addressing a much broader set of issues. In the sections below,
we demonstrate that wide-area wide-band polarimetric surveys with the SKA
will provide us with powerful new data sets aimed at addressing
the following questions:
\begin{packed_item}
\item  What is the relationship between supermassive black holes and their
environments? (\S\ref{sec_bh})
\item  What are the physical properties of absorbing systems?
(\S\ref{sec_abs})
\item How have galaxies evolved over cosmic time? (\S\ref{sec_kcorr})
\end{packed_item}
As will be explained below, \S\ref{sec_bh} and \S\ref{sec_kcorr} focus on
polarisation as a probe of the polarised sources themselves and of their
immediate environments, while \S\ref{sec_abs} considers the use of background sources to
study the effects of Faraday rotation and depolarisation in intervening
objects along the line of sight.
In \S\ref{sec_ska1} we quantify  the expected science outcomes for SKA1 and
note the relevant observing specifications. In
\S\ref{sec_ska1_early}, we consider an early science program for
SKA1 once it reaches 50\% sensitivity. In \S\ref{sec_ska2}, we
anticipate the possible science outcomes that can be pursued with SKA2.

\section{What is the Relationship Between Supermassive Black Holes and Their
Environments?}
\label{sec_bh}

All massive galaxies host supermassive black holes (SMBHs). Relativistic
outflows from these SMBHs deposit enormous amounts of mechanical energy into
their surroundings, pollute the IGM with metals and magnetic fields, and
regulate star-formation and feedback.  A key issue underpinning all these
considerations is the extent to which thermal gas from the host galaxy and
its surroundings interacts with and is entrained by these outflows. Such entrainment can
decelerate the AGN's relativistic jets, suppress star formation, regulate
the growth of the central SMBH, and control the acceleration efficiency of
ultra-high-energy cosmic rays 

Detailed polarimetric studies of individual sources have begun to reveal
some of the complex ways in which radio lobes from AGN can interact with
their environments --- for some sources we see thermal sheaths draped over
the radio lobes \citep[e.g.,][]{glc+12}, in others we observe compression
and mixing of the surrounding thermal gas \citep[e.g.,][]{glb+11}, and at
least in one case we appear to detect thermal gas {\em inside}\ the lobes
\citep{ofm+13}.

However, for the broader population we lack meaningful sample sizes,
spatially resolved spectropolarimetry, and a surrounding RM grid to correct
for foreground Galactic Faraday rotation.  With the SKA, we have the
capacity to comprehensively explore the distribution of thermal gas in and
around radio lobes from AGN.  By simultaneously measuring polarised
fractions, Faraday rotation and synchrotron intensity across a broad
bandwidth, with the high angular resolution needed to spatially
resolve lobes and distinguish them from cores and host galaxies, and in
conjunction with information on foreground and background Faraday rotation
as supplied by the RM grid, we can provide a comprehensive view of
entrainment, outflows, ionised gas and magnetic fields in radio galaxies,
covering a wide range of host galaxies, jet powers, environments and
redshifts.

\section{What Are The Physical Properties of Absorbing Systems?}
\label{sec_abs}

A fundamental limitation in radio astronomy is the difficulty in 
studying normal (star-forming) galaxies at large distances.
While the SKA will provide the  enhanced
sensitivity needed to see such sources out to higher redshifts than
currently possible,
a powerful alternative approach is to 
study otherwise invisible populations in projection against bright
background sources. For the SKA this is typified by surveys for \HI\
absorption toward distant AGN, which will allow an unbiased
study of the evolution of normal galaxies over a huge range of cosmic time
\citep{kb04,msc14}.

Polarimetry and Faraday rotation provide an equivalent probe of unseen
intervenors, but for ionised rather than neutral gas. Specifically,
the detailed magneto-ionic properties of
normal galaxies can be studied out to high redshift by determining RMs and
polarised fractions as a function of redshift, and by cross-correlating this
information with Mg\,{\sc ii}\ spectroscopy and deep optical imaging
\citep{kbm+08,bml12,bml13,focg14}.
Through this technique, we can potentially measure the amplitude of
turbulence in galactic disks and halos, the amplification time scales and
coherence lengths of galactic dynamos, and the covering fraction and spatial
extent of intervening systems, all as a function of redshift.
However, a number of technical limitations currently prevent any meaningful
progress in these areas. To advance this topic, we require:
\begin{packed_item}
\item Broad bandwidths, with which we can break the degeneracy between
different types of Faraday rotation and depolarisation along the line of
sight;
\item High sensitivity and survey speed, through which we can accumulate a
meaningful sample of sources that have both radio polarisation and optical spectroscopy data;
\item High angular resolution, with which we can
isolate the same sightlines in radio polarisation as are probed
by the corresponding optical spectra.
\end{packed_item}
The SKA fulfills all these criteria. When combined with \HI\ absorption
surveys (which perhaps can be performed commensally) plus data from
next-generation optical and infrared facilities such as 4MOST, WFIRST, LSST
and Euclid, we can obtain a complete view of the gas, turbulence,
magnetisation and spatial extents of ordinary galaxies over a wide range of
redshifts.

\section{How Have Galaxies Evolved Over Cosmic Time?}
\label{sec_kcorr}

\begin{figure}[t!]
\centerline{\psfig{file=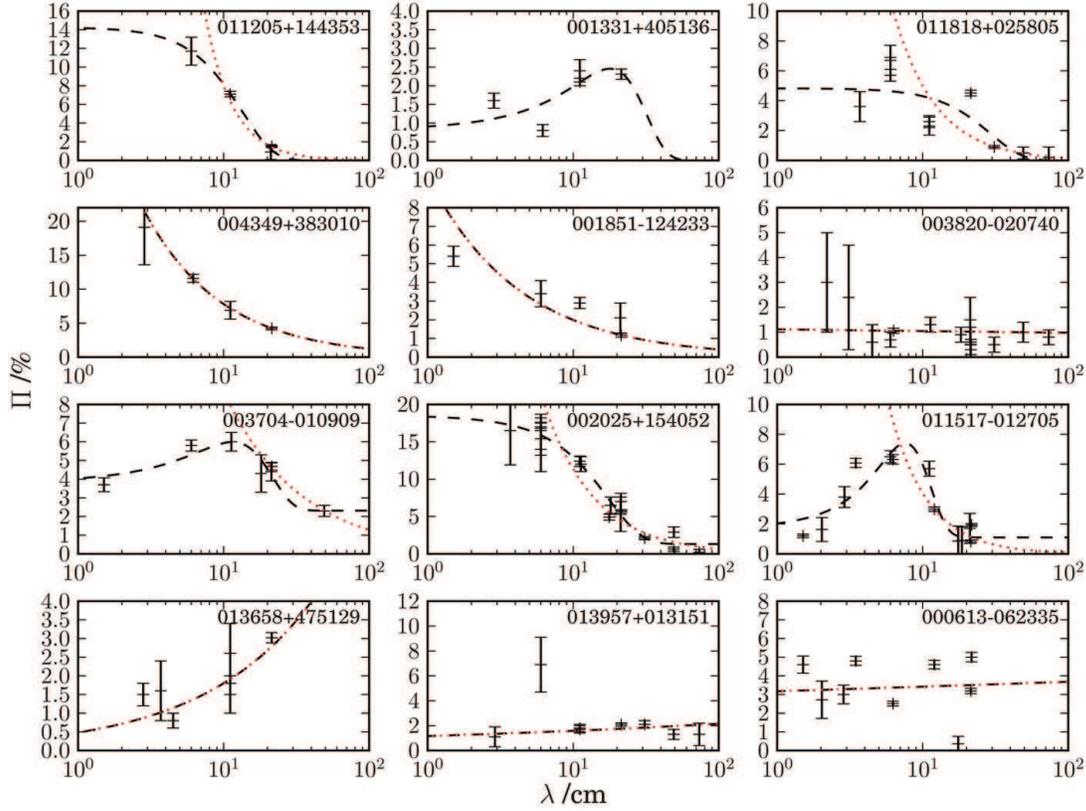,width=\textwidth}}
\caption{$\Pi$ vs $\lambda$
for 12 sources, reproduced from \protect\cite{fgc14} by
permission of the AAS. The
best-fit polarisation SED chosen from four simple models is shown as a black
dashed line: the top row shows sources whose best-fitting polarisation SED
has a Gaussian profile, the second row shows depolarising sources, the third
row shows SEDs modelled by a Gaussian with a constant term and the last row
shows repolarising sources. The red dotted line shows the best-fit power law
for $\Pi$ vs $\lambda$. The quality of the fits decreases from left to
right.}
\label{fig_farnes}
\end{figure}

The experiment described in
\S\ref{sec_bh} is one of many measurements in which we wish to study the
intrinsic properties of radio sources, and to determine how radio galaxies
and their associated SMBHs and host galaxies have evolved as a function of
redshift.
However, any comparison of such sources at different redshifts
will be meaningless unless
the observed properties are first corrected into the frame of reference in
which the emission or Faraday rotation occurs. Usually this
requires the application of a ``K-correction'' \citep[e.g.,][]{hbbe12}, in which the data
are shifted to a higher frequency than that with which they were observed.
For many applications,
the K-correction is simply a direct extrapolation of the source's observed flux
using the redshift and spectral index. However, for polarimetry the
situation is more complex. Polarisation from AGN as a function of
wavelength can show a wide variety of
behaviours: some sources depolarise, others ``repolarise'', and others show
oscillatory or other behaviour, as shown in Figure~\ref{fig_farnes}.  These phenomena are further
compounded when a polarised source is spatially unresolved, and contains
multiple components with different spectral indices and different polarised
fractions. 

Thus to determine the polarisation properties of a source in its rest frame,
one must directly observe the emitted and subsequently redshifted radiation
--- extrapolation to frequencies outside the observing 
band is generally not possible.
Such studies require near-continuous polarisation data observed or interpolated  across a wide range of
frequencies. The wide band is needed both to compare sources 
over a range of redshifts at a single
fixed emitting wavelength, and to characterise the
polarisation spectral energy distributions (SEDs) of individual sources.
\cite{fgc14} have presented the first
large catalogue of polarisation SEDs suitable for comparison of objects
at different redshifts in a common rest frame,
consisting of 951 sources with multiple polarisation
measurements over the frequency range 0.4--100~GHz (see examples in
Fig.~\ref{fig_farnes}). While the frequency range covered is very large, the
sampling is sparse and inconsistent, the individual data points come from
different surveys taken at different epochs with different telescopes and
differing angular resolutions, and the overall sample size is small. 

Using the SKA, broadband polarimetry can provide a very large sample of
polarisation SEDs, suitable for robust and accurate rest-frame corrections
for many different types of sources over a wide range of
redshifts.\footnote{Note that two sources at different redshifts will have
different effective spacings of adjacent frequency channels in their rest
frames. This will mean that a large RM seen at one redshift might not be
detectable at another redshift, as per Equation~(63) of \cite{bd05}.
However, in practice the expected channel widths for the SKA will be
sufficiently narrow that all physically plausible values of RM will be
detectable at all redshifts.} This will enable a diverse set of powerful
experiments with which we can study the polarisation properties of radio
galaxies and AGN as a function of redshift. In addition, the SKA will
simultaneously obtain total intensity SEDs for an even larger number of
sources, which will facilitate detailed studies of spectral indices,
spectral curvature and spectral turnover.

\section{Science Outcomes for SKA1}
\label{sec_ska1}

The experiments proposed above all require sensitive high-resolution
polarisation observations of a large sample of sources over a broad
bandwidth. Both SKA1-MID and SKA1-SUR meet these basic specifications.
For the RM grid experiment, two additional requirements were all-sky
coverage (to fully probe the magnetic field of different parts of the Milky
Way) and a very dense sampling of polarised sources (to provide multiple
closely spaced sightlines through extended foreground sources).  The first
of these considerations is not relevant for our purposes, and the second is
of reduced importance. Instead, our main drivers are to optimise the
frequency coverage and to maximise the number of sources detected. 
A full consideration of all survey specifications is beyond the scope of
this Chapter, but
as for other proposed polarisation and continuum surveys, consideration
needs to be given to the requirements on dynamic range, flux calibration
(both relative and absolute), $u-v$ coverage, mosaicing strategy 
and wide-field polarisation purity.

\subsection{Choice of Observing Frequency}
\label{sec_freq}

\cite{ab11} have considered
the optimum observing frequency, \textnu$_{\rm opt}$, for the study of magnetic fields. 
They define \textnu$_{\rm opt}$ as the frequency at which the polarised intensity is a
maximum, while requiring a broad bandwidth around this
frequency to distinguish between different Faraday and depolarisation
effects. Within a given source or intervenor population, at a redshift
$z$ and for a standard
deviation \textsigma$_{\rm RM}$ in RM or Faraday depth (in the frame of the Faraday
rotating medium),
we have that \textnu$_{\rm opt} \propto$~\textsigma$_{\rm RM}^{1/2}~(1+z)^{-1}$. 
For the case of entrained thermal gas in and around the lobes of radio
galaxies as considered in \S\ref{sec_bh}, we adopt \textsigma$_{\rm RM}
\sim 3-6$~rad~m$^{-2}$ \citep{ofm+13}, for which we infer \textnu$_{\rm opt}
\approx 900-1300~(1+z)^{-1}$~MHz. For intervening galaxies along the line of
sight as discussed in \S\ref{sec_abs}, we adopt the ``faint galaxy halo'' 
case of \cite{ab11}, for which $|{\rm RM}| \approx$~\textsigma$_{\rm RM} \approx
8$~rad~m$^{-2}$ and \textnu$_{\rm opt} \approx 1200-1500~(1+z)^{-1}$~MHz. 
Finally, for corrections for redshift as proposed in \S\ref{sec_kcorr}, we wish to
maximise the redshift range over which sources within a given population
can be compared. To achieve this, we must
observe over as broad a band as possible,
encompassing the optimal
frequency \textnu$_{\rm opt}$.
SKA1-MID Band~2 (950--1760~MHz)
and SKA1-SUR Band~2 (650--1670~MHz) both provide the required frequency
coverage.
SKA1-SUR is preferred because it extends down to lower
frequencies than SKA1-MID: in $\lambda^2$-space, the coverage of Band~2 is
2.5 times larger for SKA1-SUR than for SKA1-MID.  The wavelength coverage
of SKA1-MID could be extended by additionally observing with SKA1-MID Band~1
(350--1050~MHz), but the total observing time required would be larger than
for SKA1-SUR, despite the larger instantaneous bandwidth for SKA1-MID.

{\em In subsequent discussion, we consider a broadband continuum
polarisation survey using SKA1-SUR~2 (650--1670~MHz).}\footnote{We note that
a possible shift of SKA1-SUR Band~2 to 560--1430~MHz is currently under
consideration.}
This optimises the study of radio
lobes (\S\ref{sec_bh}) for emitting sources in the redshift range  $z \la
0.4-1$ and for intervenors
(\S\ref{sec_abs}) in the range $z \la 0.8 -1.3$, while allowing
corrections into a common rest frame
(\S\ref{sec_kcorr}) for a population covering the redshift range $\Delta z
\approx 1.6(1+z_{\rm min})$ for a minimum redshift $z_{\rm min}$. Thus all
three topics can be simultaneously pursued for sources or intervenors 
in the redshift range $0<z<1$.

\subsection{Choice of Angular Resolution}

The experiments described above require angular resolution sufficiently high
to spatially resolve the polarised emission from radio lobes
(\S\ref{sec_bh}), to match polarisation information to the locations of
optical spectra (\S\ref{sec_abs}), and to resolve the polarised emission
into multiple components (\S\ref{sec_kcorr}).  Angular resolutions of
$\sim$$10''$ to $\sim$$1'$,
while suitable for the RM grid, cannot address the questions
envisaged here.

Our knowledge of polarised morphologies for the faint radio sky is limited.
The two deepest polarimetric observations are those of \cite{hngm14,hng+14}
and \cite{ro14}. \cite{hngm14,hng+14} presented observations 
at $10''$ resolution, covering the frequency range 1.3--1.5~GHz
to a sensitivity of $\approx30\mu$Jy~beam$^{-1}$. \cite{ro14} have reported
observations at $1\farcs6$ resolution, covering
the frequency range 1.3--1.9~GHz to a sensitivity of
$\approx2.4$~$\mu$Jy~beam$^{-1}$. At $10''$ resolution, 36\% of polarised
sources show extended structure or multiple components in polarised
intensity \citep{hngm14}, while at $1\farcs6$ resolution, 69\% of polarised
sources show spatial extent in polarisation\footnote{Almost all these extended polarised
sources fall in the redshift range $z\sim 0.3-1$, matching the redshift
range covered by SKA1-SUR Band~2 as discussed in \S\ref{sec_freq} above}
\citep{ro14}.  Thus
arcsecond-level (or better) resolution is desired, so as to maximise the
number of extended polarised sources detected. This choice also provides the
angular resolution needed to compare sources to optical data, and avoids
confusion in total intensity even for relatively long integrations. In
summary, the angular resolution of the full SKA1-SUR array ($1''$ at
1.4~GHz) is a good match to our requirements.

\subsection{Survey Specifications}
\label{sec_specs}

We first consider the science requirements to study the interaction of radio
lobes with thermal gas
(\S\ref{sec_bh}). To fully investigate this phenomenon,
we need a sample large enough to group such
sources into 10 redshift bins, 10 luminosity bins and 10 morphological
categories \citep[see \S6.2.2 of][]{hng+14}, with 100 sources in each of
these 1000 subsets. We thus aim to observe a total of $10^5$ polarised,
extended, radio galaxies.  Assuming that $\sim70\%$ of polarised sources are
spatially extended at arcsec-resolution \citep{ro14} and that $\sim20\%$ of
sources show complex Faraday depth spectra that might be the signature of
interactions with
thermal gas, we must observe $7\times10^5$ polarised sources to
obtain our required data set, or a sky density of $\sim23$ sources/deg$^2$
for a survey covering 30\,000~deg$^2$. 
This sky density is reached
for a peak polarised intensity (at 1.4~GHz and at $1\farcs6$ resolution) of
$\approx0.1$~mJy~beam$^{-1}$ \citep{ro14}.  The signal-to-noise ratio required for
analyses such as that shown in Figure~\ref{fig_obr12} is a topic under
active investigation \citep{sra+14}, but preliminary results suggest that a
20\textsigma\ detection is sufficient.  We therefore require an rms sensitivity
of 5~$\mu$Jy~beam$^{-1}$ over the entire bandpass (650--1670~MHz). To cover
this bandpass with a 500-MHz instantaneous bandwidth, we require two passes
over the sky: a survey at 900~MHz with a 55\% fractional bandwidth, and a
survey at 1420~MHz with a 35\% fractional bandwidth. For a typical source
spectral index $\alpha = -0.7$, this requires an rms of
8.5~$\mu$Jy~beam$^{-1}$ at 900~MHz, and 6.2~$\mu$Jy~beam$^{-1}$ at 1420~MHz.
Figure~9 of \cite{bra14} presents the sensitivity reached by SKA1-SUR as
a function of frequency and resolution, assuming a 30\% fractional bandwidth
and 10~hours per pointing.  Scaling from these numbers at $1''$, we require
4.4 and 1.3~hrs per pointing at 900 and 1420~MHz respectively.  For
$\approx1700$ pointings over 30\,000~deg$^2$, the total survey time is
around 10\,000 hours. We note that the 1420-MHz component of this survey
will simultaneously provide the data needed for observations of an RM grid
\citep{joh14}.

For the case of intervenors along the line of sight (\S\ref{sec_abs}), we
motivate our survey on the results of \cite{focg14}, who were able to show a
difference in the radio polarisation properties of quasars with and without
foreground Mg\,{\sc ii}\ absorbers at 3.5\textsigma\ significance,
using $\sim$140 quasars of which $\sim$40\% have one or more Mg\,{\sc ii}\ absorbing
systems along the line of sight.  For SKA1, we aim to double this sample
size in subcategories divided into 10 bins of redshift, 5 bins for absorber
equivalent-width, and 5 bins of intervenor impact parameter, for a total of
50\,000 quasar sightlines.  Estimating that quasars are $\sim10\%$ of the
radio source population at the relevant flux levels in total intensity, we
therefore require $\sim7\times10^5$ polarised sources in total,
or again a sky density of 23 sources/deg$^2$
over 30\,000 deg$^2$. 
The required rms can be achieved by the same survey as described above.
Note that we assume that all 70\,000 radio-loud quasars will have associated
optical spectroscopy from future surveys, which is reasonable given that the
required sky density of such spectra ($\la$2~sources/deg$^2$) is well below
that already available in the northern hemisphere through SDSS ($\ga10$
quasar spectra per deg$^2$).

The total yield of the proposed survey will be $7\times10^5$ polarised
sources detected at $\ge$20-\textsigma\ significance, corresponding to total
intensity fluxes $\ga5-10$~mJy. 
We will be able to make use of $\sim$75\% of radio
galaxies at these flux levels \citep{wmj+08} if we seek to apply
rest-frame corrections for redshifts $0.5 \la z \la 3$
(see~\S\ref{sec_freq}).

\section{An Early Science Program for SKA1}
\label{sec_ska1_early}

A valuable early science program on broadband polarimetry can be pursued
once SKA1 reaches 50\% of its full sensitivity. At this stage, SKA1-SUR
will consist of 36 12-metre dishes and 24 15-metre dishes, and will have
$\sim$5 times the survey speed of ASKAP. 
One of the survey programs intended for ASKAP is the Polarisation Sky Survey
of the Universe's Magnetism \citep[POSSUM;][]{glt10}, which will image
30\,000~deg$^2$ of the polarised sky to a sensitivity of
$\sim10$~$\mu$Jy~beam$^{-1}$ at 10$''$ resolution and over
the frequency range 1.2--1.5~GHz.  Early science for SKA1 could consist of a
survey with similar sky coverage, angular resolution and sensitivity
to that planned for POSSUM, but covering the frequency range 700--1200 MHz.
This would require 2--3 months of observing time with the first 50\% of
SKA1-SUR. The two data sets would then be combined to provide
700--1500~MHz observations at $\approx$7~$\mu$Jy~beam$^{-1}$ sensitivity
at $10''$ resolution.

While such a data set would not have the high angular
resolution and large frequency coverage offered by the  full set of broadband
observations described in \S\ref{sec_ska1}, it would provide vital input
into the design and approach of these subsequent programs. For example, for
studying the interaction of radio lobes with their environments
(\S\ref{sec_bh}), the proposed early science program would identify
$\sim10^4$ polarised radio lobes that would be spatially resolved at
$\approx10''$ resolution and that would show signatures of thermal
entrainment. This is 10\% of the full sample size envisaged in
\S\ref{sec_specs}, which would allow us to identify and study the most
prominent cases, and to develop the algorithms and models needed to
interpret the larger data sets to follow.

\section{Considerations for SKA2}
\label{sec_ska2}

The increased sensitivity of SKA2 will obviously allow measurements of
effects due to much smaller magnetic field strengths and electron densities
in a much larger sample of sources.  However, the key point for 
polarimetry experiments with SKA2 is the increase in the available frequency
range, i.e., wide-field survey capability covering 350--1500~MHz.  This will
allow an extension of the studies described in \S\ref{sec_freq} out to
redshifts $z \sim 3-4$, corresponding to lookback times when we expect
substantial evolution in magnetic field geometry and dynamo activity
\citep{abks09}. Over the next 5 years, new results on the polarised sky from
POSSUM, WODAN and the JVLA will greatly advance our understanding of all the
issues discussed throughout this Chapter, and will help fully define the
goals and expected outcomes for SKA2.

\section{Summary and Conclusions}

We have described a new polarimetric landscape in which a wide range of
Faraday effects have now been identified, each of which provides a unique
probe of fundamental properties of galaxies. Over relatively
narrow fractional bandwidths ($\la30\%$), these effects are either
undetectable or
highly degenerate. However, by using sensitive
spectropolarimetric surveys over broad bandwidths, we can reveal the
mechanisms by which supermassive black holes couple to their environments,
can measure the properties of ordinary galaxies over a large range in
redshift, and can understand how different radio source populations have
evolved over the Universe's history. 
Using SKA1-SUR over its full Band~2 frequency range (650--1670 MHz), we can
detect polarised emission from the tens of thousands of sources needed to
address all these experiments out to redshifts beyond $z\sim1$. With SKA2,
the increased sensitivity and broader frequency coverage will
allow us to extend these observations to redshifts $z \sim3-4$, allowing us
to couple existing studies of galaxy evolution and star-formation history to
as yet unaddressed questions of how turbulence, ionisation fraction,
entrainment efficiency  and magnetic field properties have all evolved over
cosmic time.

\acknowledgments 

B.M.G. acknowledges the support of the Australian Research Council (ARC) through
an Australian Laureate Fellowship (FL100100114), and through the ARC Centre
of Excellence for All-sky Astrophysics (CE110001020). We thank the referee
for a constructive review of the original manuscript.

\bibliographystyle{apj_5authors_etal}
\bibliography{journals,modrefs,psrrefs,crossrefs}

\end{document}